\documentclass[twocolumn,aps,showpacs]{revtex4}

\def  \be   {\begin{equation}}
\def  \ee   {\end{equation}}
\def  \beq  {\begin{eqnarray}}
\def  \eeq  {\end{eqnarray}}

\begin{document}

\title{On the two-dimensional magnetic reconnection with nonuniform resistivity}

\author{Leonid M.~Malyshkin}
\email{leonmal@flash.uchicago.edu}
\affiliation{Department of Astronomy, The University of Chicago --
The Center for Magnetic Self-Organization (CMSO), Chicago, IL 60637.
}

\author{Russell M.~Kulsrud}
\email{rmk@pppl.gov}
\affiliation{
Princeton Plasma Physics Laboratory -- The Center for 
Magnetic Self-Organization (CMSO), Princeton, NJ 08543. 
}

\date{\today}

\begin{abstract}
In this paper two theoretical approaches for the calculation of the rate of
quasi-stationary, two-dimensional magnetic reconnection with nonuniform
anomalous resistivity are considered in the framework of incompressible 
magnetohydrodynamics (MHD). In the first, ``global'' equations approach 
the MHD equations are approximately solved for a whole reconnection 
layer, including the upstream and downstream regions and the layer center. 
In the second, ``local'' equations approach the equations are solved 
across the reconnection layer, including only the upstream region and the 
layer center. Both approaches give the same approximate answer for the 
reconnection rate. Our theoretical model is in agreement with the results 
of recent simulations of reconnection with spatially nonuniform 
resistivity by Baty, Priest and Forbes (2006), contrary to their 
conclusions.
\end{abstract}


\maketitle


\section{Introduction}
\label{INTRODUCTION}

Magnetic reconnection is one of the most important processes of plasma
physics, and is believed to play the central role in observed phenomena 
in laboratory and cosmic plasmas. At the same time there has been a 
long-standing debate about the correct theoretical model of magnetic 
reconnection. Most previous theoretical and numerical work focused on
reconnection processes in two-dimensions, in which all physical scalars and 
vectors are independent of the third coordinate ($z$). There exist two original
and well known models of magnetic reconnection with constant resistivity. 
First, the Sweet-Parker reconnection model~\cite{sweet_1958,parker_1963}, 
which predicts a slow magnetic reconnection rate in hot low-density plasmas.
Second, the Petschek model~\cite{petschek_1964}, in which a fast 
reconnection rate is achieved by introducing switch-off magnetohydrodynamic 
(MHD) shocks attached to the ends of a relatively short reconnection layer 
in the downstream regions. Many numerical simulations have been carried out 
to discriminate between these two models. More recent high-resolution
simulations generally favor the Sweet-Parker model of slow reconnection
in the case of constant resistivity, and do not confirm the Petschek 
theoretical picture for the geometry of the reconnection layer with 
shocks~\cite{biskamp_1986,breslau_2003,uzdensky_2000}. However, reconnection
becomes much faster and Petschek-like if resistivity is not constant and
is enhanced locally in the reconnection layer (numerical studies of this
case were pioneered by Ugai and Tsuda~\cite{ugai_1977,tsuda_1977}, by Hayashi
and Sato~\cite{hayashi_1978,sato_1979}, and by Scholer~\cite{scholer_1989}).

These results have been called into question in a recent paper,
Baty, Priest and Forbes (2006)~\cite{baty_2006}, which claims that Petschek
reconnection can occur when the resistivity is not absolutely constant but 
varies by an arbitrarily small amount. This paper has motivated us show on 
the basis of an earlier paper of ours, Malyshkin 
{\it et~al$.$}~\cite{malyshkin_2005}, that the Petschek shocks are produced not 
by the variation of the resistivity but by the rate of variation which for 
their case is very large and unlikely to happen naturally.

In 2001 Kulsrud~\cite{kulsrud_2001} provided some theoretical insight 
into magnetic reconnection process which qualitatively explained the results
of recent simulations. He considered quasi-stationary, two-dimensional
magnetic reconnection in the classical Sweet-Parker-Petschek reconnection
layer with zero guide field ($B_z=0$), zero plasma viscosity and an anomalous
resistivity that is a piecewise linear function of the electric current.
Kulsrud was first to suggest that one has to calculate the 
half-length of the reconnection layer $L'$ from the MHD equations and 
the jump conditions on the Petschek shocks, instead of treating $L'$ 
as a free parameter (as Petschek erroneously did when chose $L'$ to be 
equal to its minimal possible value under the condition of no 
significant disruption to the plasma flow). As a result, in the case
of constant resistivity Kulsrud correctly estimated the layer half-length 
$L'$ to be approximately equal to the global magnetic field scale, 
$L'\approx L$. In this case the Petschek reconnection rate reduces to the 
slow Sweet-Parker reconnection rate~\cite{kulsrud_2001,kulsrud_2005}, in
agreement with numerical simulations. The second result obtained
by Kulsrud is that in the case when resistivity is non-constant but 
anomalous and enhanced (e.g.~by plasma instabilities), the reconnection 
rate can become considerably faster than the Sweet-Parker rate.

In our recent paper (Malyshkin, Linde and Kulsrud~\cite{malyshkin_2005}),
to which we will hereafter refer as MLK2005 paper, we put Kulsrud's 
derivations on a rigorous analytical basis and extended his model by 
using a new theoretical approach to calculate the reconnection 
rate. This approach is based on ``local'' analytical derivations across 
a thin reconnection layer, and it is applicable to the case when 
resistivity is anomalous, nonuniform and an arbitrary function 
of the electric current and of the spatial coordinates. We included the 
case of non-zero guide field $B_z\ne0$ and non-zero plasma viscosity 
in our model. We found an approximate formula for the reconnection 
rate which confirmed Kulsrud's theoretical results. 

The present paper has two goals. First, to calculate the reconnection
rate, and second, to compare it with the recent simulations of Baty 
{\it et~al$.$}~\cite{baty_2006}.

In the next section we simultaneously follow two theoretical approaches 
to the calculation of the reconnection rate. In the first approach, 
which we call the ``global'' equations approach, we derive and solve 
approximate MHD equations for a whole reconnection layer, including 
the upstream and downstream regions and the layer center. These 
theoretical derivations are similar to those done 
before, except for an important difference. Namely, we take into 
consideration an additional important equation, the spatial
homogeneity of the z-component of electric field $E_z$ along the 
reconnection layer. This equation, together with the jump condition on 
Petschek shocks, allows us to find the reconnection layer length $L'$, 
which must be determined consistently~\cite{kulsrud_2001}. The second 
theoretical approach, which we call the ``local'' equations approach, 
basically coincides with the calculations done in the MLK2005 paper. 
In this approach we derive and solve approximate MHD equations across
a thin reconnection layer, including only the upstream region and the 
layer center. Both approaches give the same approximate formula for 
the reconnection rate, which is valid in the general case of an
arbitrary nonuniform anomalous resistivity [see Eq.~(\ref{RATE})]. 
Using these two approaches simultaneously allows us to better 
understand the physics of a reconnection process.

In Sec.~\ref{LOCALIZED_RESISTIVITY} we pursue the second goal of this 
paper. We consider the case treated by Baty {\it et~al$.$}~\cite{baty_2006}, 
when resistivity is a prescribed function of the two spatial coordinates 
$x$ and $y$ (remember that we consider two-dimensional reconnection, so 
that no physical quantities depend on $z$). We argue that our theoretical
model is in agreement with the results of these recent simulations of 
reconnection by Baty, Priest and Forbes~\cite{baty_2006}. This agreement
contradicts their theoretical conclusion that Petschek shocks exist 
with constant resistivity, and we explain why. Finally, in 
Sec.~\ref{DISCUSSION} we discuss our results.


\section{Reconnection with nonuniform anomalous resistivity}
\label{ANOMALOUS_RESISTIVITY}

\begin{figure}[t]
\vspace{3.6truecm}
\includegraphics{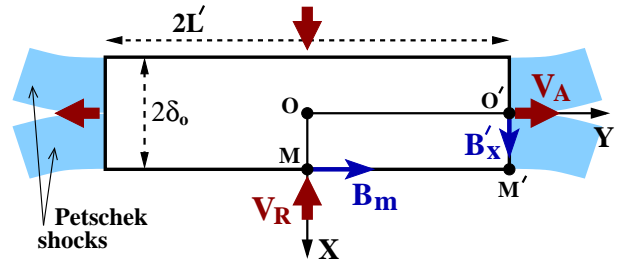}
\caption{(Color online) The geometrical configuration of the classical
Sweet-Parker-Petschek reconnection layer is shown. Petschek shocks exist only 
if the reconnection is considerably faster than the Sweet-Parker reconnection 
rate.
} 
\label{FIGURE_LAYER}
\end{figure}

In this section we consider magnetic reconnection with nonuniform 
anomalous resistivity in the classical two-dimensional 
Sweet-Parker-Petschek reconnection layer, shown in 
Fig.~\ref{FIGURE_LAYER}. The layer is in the x-y plane with the 
x- and y-axes being perpendicular to and along the layer 
respectively. The length of the layer is 
$2L'$, which is approximately equal to or smaller than the global 
magnetic field scale $L$ that will be introduced below. The 
thickness of the layer, $2\delta_o$, is much smaller 
than its length, i.e.~$2\delta_o\ll2L'$. The classical 
Sweet-Parker-Petschek reconnection layer is assumed to have a 
point symmetry with respect to its geometric center point~$O$ in
Fig.~\ref{FIGURE_LAYER} and reflection symmetries with 
respect to the axes $x$ and $y$.
Thus, for example, the x- and y-components of the plasma velocities 
${\bf V}$ and of the magnetic field ${\bf B}$ have the following
simple symmetries:  
$V_x(\pm x,\mp y)=\pm V_x(x,y)$, $V_y(\pm x,\mp y)=\mp V_y(x,y)$, 
$B_x(\pm x,\mp y)=\mp B_x(x,y)$ and $B_y(\pm x,\mp y)=\pm B_y(x,y)$.
There could be a pair of Petschek shocks attached to each of the 
two reconnection layer ends in the downstream regions. Because 
of the MHD jump conditions on the Petschek shocks, there must 
be a nonzero perpendicular magnetic field $B_x'$ in the 
downstream region at point~$O'$ in 
Fig.~\ref{FIGURE_LAYER}~\cite{kulsrud_2001,kulsrud_2005}. 
If the plasma viscosity is small, then the plasma outflow velocity
$V_{\rm out}'$ in the downstream region at point~$O'$ is approximately 
equal to the Alfven velocity $V_A$ calculated in the upstream region 
at point~$M$ (refer to Fig.~\ref{FIGURE_LAYER}). The 
plasma inflow velocity $V_R$ in the upstream region at point~$M$,
outside the reconnection layer, is much smaller than the 
outflow velocity, $V_R\ll V_A$. Finally, the magnetic field $B_m$ 
at point~$M$, outside the layer, is mostly in the direction of 
the layer (i.e.~in the y-direction). 

Now let us list four assumptions that we make about reconnection 
process. First, we assume that the characteristic Lundquist number 
is large, which (by our definition) means that resistivity is 
negligible outside the reconnection layer. Second, we assume the 
plasma flow is incompressible. Third, for simplicity we neglect plasma 
viscosity. (The case of nonzero viscosity is treated in the MLK2005 
paper~\cite{malyshkin_2005}.) Fourth, we assume the reconnection 
process is quasi-stationary. This is true if the reconnection is 
slow, $V_R/V_A\ll1$, and that there are no plasma instabilities in 
the reconnection layer. For an incompressible viscousless plasma the 
assumption of slow reconnection is equivalent to the assumption
that the reconnection layer is thin, $\delta_o/L'\approx V_R/V_A\ll1$. 
The above assumptions are standard in the Sweet-Parker and Petschek 
reconnection models. Please note that we make no assumptions
about the values of the guide field $B_z$. In other words, our 
derivations apply to what is called ``two-and-a-half dimensional''
reconnection.

For brevity, we use physical units in which the speed of light and 
four times $\pi$ are replaced by unity, $c=1$ and 
$4\pi=1$~\footnote{Equations in the CGS units can be obtained by 
substitutions 
${\bf B}\rightarrow {\bf B}/\sqrt{4\pi}$ for magnetic field, 
${\bf E}\rightarrow c{\bf E}/\sqrt{4\pi}$ for electric field, 
${\bf j}\rightarrow (\sqrt{4\pi}/c){\bf j}$ for electric current, 
and $\eta\rightarrow\eta$ for resistivity.
}.
In these units the MHD equations that we need to find the reconnection 
rate are as follows. Faraday's law 
${{\bf\nabla}\times{\bf E}}=-\partial{\bf B}/\partial t$ for 
the x- and y-components of magnetic field in two dimensions is
\beq
\partial E_z/\partial y=-\partial B_x/\partial t\approx 0,
\quad
\partial E_z/\partial x=\partial B_y/\partial t\approx 0,
\label{FARADAYS_LAW_EXACT}
\eeq
where $\partial{\bf B}/\partial t\approx 0$ because of the
quasi-stationarity of reconnection. From Eqs.~(\ref{FARADAYS_LAW_EXACT})
we see that the electric field z-component is constant 
in space, i.e.~$E_z=E_z(t)$ is a function of time only.
Next, neglecting the displacement current in the framework of MHD, 
Ampere's law for the z-component of the current is
\beq
j_z=({\bf\nabla}\times{\bf B})_z=\partial B_y/\partial x-\partial B_x/\partial y.
\label{AMPERES_LAW_EXACT}
\eeq
Ohm's law for the spatially uniform z-component of the electric 
field is
\beq
E_z(t)=-V_xB_y+V_yB_x+\eta j_z=\mbox{constant in space},
\label{OHMS_LAW_EXACT}
\eeq
where resistivity $\eta=\eta(j_z,x,y)$ is an arbitrary function 
of the electric current z-component and of the two-dimensional 
coordinates~\footnote{
We assume that $\eta(j_z,x,y)$ has finite derivatives in $y$ up to 
the second order and in $x$ and $j_z$ up to the first order. 
We consider $\eta$ to be a function of $j_z$ instead of the total
current $j=(j_z^2+j_x^2+j_y^2)^{1/2}$. This is because the electrical 
conductivity in the z-direction can be reduced by plasma instabilities 
due to large values of $j_z$.
}.
Next, the equation for plasma acceleration along the layer in the y-direction is
\beq
\rho({\bf V}\cdot{\bf\nabla})V_y=-(\partial/\partial y)[P+B_x^2/2+B_y^2/2]
+({\bf B\nabla})B_y,
\label{ACCELERATION_EXACT}
\eeq
where $\rho$ is the constant plasma density, $P$ is the sum of 
the plasma pressure and the guide field pressure $B_z^2/2$. 
In addition, we have
\beq
\partial_x V_x+\partial_y V_y=0,\quad
\partial_x B_x+\partial_y B_y=0
\label{DIVERGENCE_EXACT}
\eeq
because the field and the velocity are divergence-free. 

Now, from MHD equations~(\ref{AMPERES_LAW_EXACT})--(\ref{DIVERGENCE_EXACT})
we obtain the following table of ``global'' and ``local'' equations for
the reconnection layer:
\begin{widetext}
\vspace{-4.0ex}
\beq
\begin{tabular}{|cl|c|c|}
\hline
 & & {\bf Global Equations} & {\bf Local Equations} \\
\hline
$1^{\rm st},$ & Ampere's law & 
\multicolumn{2}{c|}{$j_o\equiv(j_z)_o\approx
(\partial_x B_y)_o\approx B_m/\delta_o$} \\
\hline
$2^{\rm nd},$ & Incompressibility &
$V_RL'\approx V_{\rm out}'\delta_o$ &
$(\partial_y V_y)_o=-(\partial_x V_x)_o\approx V_R/\delta_o$ \\
\hline
$3^{\rm rd},$ & Plasma acceleration &
$V_{\rm out}'\approx V_A\equiv B_m/\sqrt{\rho}$ &
$\rho{(\partial_y V_y)_o}^2=-(\partial_y^2P)_o+
j_o(\partial_y B_x)_o\vphantom{\Big|}$ \\
\hline
$4^{\rm th},$ & Jump condition on shocks & $B_x'/\sqrt{\rho}=V_R'\approx V_R$ & \\
\hline
$5^{\rm th},$ & $E_z={\rm const}$ across the layer &
\multicolumn{2}{c|}{$\eta_oj_o=V_RB_m$} \\
\hline
$6^{\rm th},$ & $E_z={\rm const}$ along the layer &
$\eta_oj_o=\eta'j_z'+V_{\rm out}'B_x'$ &
$0=\partial_y^2\big(\eta j_z\big)_o+
2(\partial_y V_y)_o(\partial_y B_x)_o\vphantom{\Big|}$ \\
\hline
$7^{\rm th},$ & Unknown quantities & $j_o$, $\delta_o$, $V_R$, $L'$, $V_{\rm out}'$, $B_x'$ & 
$j_o$, $\delta_o$, $V_R$, $(\partial_y V_y)_o$, $(\partial_y B_x)_o$\\
\hline
\end{tabular}
\label{TABLE}
\eeq
\end{widetext}
Here quantities with subscript $_o$ are taken at the reconnection layer central 
point~$O$, while quantities with the prime sign are taken at point $O'$
in the downflow region (refer to Fig.~\ref{FIGURE_LAYER}). The column 
``Global Equations'' includes equations that are written at points~$O$,~$M$ 
and~$O'$, these equations represent the ``global'' equations approach to
the calculation of the reconnection rate. The column ``Local Equations''
includes equations that are written only at points~$O$ and $M$ and represent 
the ``local'' equations approach. Note that equations on lines~1 and~5 enter 
both columns in the same form. The first line of the table includes the Ampere's 
law equation~(\ref{AMPERES_LAW_EXACT}), with the $\partial_y B_x$ term 
neglected because it is small, and the $\partial_x B_y$ term estimated 
at point~$O$. The second line of the table includes the plasma incompressibility 
condition~(\ref{DIVERGENCE_EXACT}) in its ``global'' and ``local'' forms. The 
``global'' form is the mass conservation equation, while in the ``local'' form the 
$\partial_x V_x$ term is estimated at point~$O$. The third line contains 
equations for plasma acceleration given by Eq.~(\ref{ACCELERATION_EXACT}).
The ``global'' equation $V_{\rm out}'\approx V_A$ reflects 
the well known result that in the absence of viscosity the plasma outflow 
velocity in the downstream region is approximately equal to the Alfven velocity 
calculated in the upstream region~\cite{sweet_1958,parker_1963,petschek_1964}. 
The ``local'' equation for plasma acceleration results from differentiation of 
Eq.~(\ref{ACCELERATION_EXACT}) with respect to $y$ and taking the 
symmetries of the reconnection layer into account. The $j_o(\partial_y B_x)_o$ 
term in this equation is the magnetic tension force, while the pressure term is
equal to 
\beq
(\partial_y^2 P)_o =\partial_y^2\big(B_y^2/2\big)_m=B_m(\partial_y^2 B_y)_m<0.
\label{PRESSURE_TERM}
\eeq
Thus, the drop of pressure $P$ (which includes the plasma and guide field 
pressure) along the layer is equal to the pressure drop of the parallel 
magnetic field outside the layer. This result follows from the force balance 
condition for the plasma across the reconnection layer (in analogy with the 
Sweet-Parker derivations), and its rigorous proof can be found in the 
appendix~A of the MLK2005 paper~\cite{malyshkin_2005}. The last equality in 
Eq.~(\ref{PRESSURE_TERM}) comes from the layer reflection symmetry with 
respect to the $x$ axis. 

Next, the fourth line in Table~\ref{TABLE} includes the standard  
jump condition on the switch-off MHD Petschek shocks attached to the ends of 
the reconnection layer~\cite{kulsrud_2001,kulsrud_2005}. This condition is 
a ``global'' equation, and the plasma incoming velocity $V_R'$ at point~$M'$ 
is estimated as being approximately equal to the plasma incoming velocity 
$V_R$ at point~$M$ (refer to Fig.~\ref{FIGURE_LAYER}). There is no 
corresponding ``local'' equation because we do not consider the downstream 
region and Petschek shocks in the ``local'' equations approach!  
Equations in the fifth and sixth lines of Table~\ref{TABLE} directly 
result from Ohm's law equation~(\ref{OHMS_LAW_EXACT}). Namely, we use 
the spatial homogeneity of $E_z(t)$, which is a consequence of the 
quasi-stationarity of reconnection. To obtain the single equation in the 
fifth line and the ``global'' equation in the sixth line, we equate the 
Ohm's law expression for $E_z$ at points~$O$,~$M$ and~$O'$. 
To obtain the ``local'' equation in the sixth line, we take the second order 
partial derivative $\partial_y^2$ of Eq.~(\ref{OHMS_LAW_EXACT}) at point~$O$.
Finally, the seventh line in Table~\ref{TABLE} lists all unknown physical 
quantities to be estimated in the ``global'' and ``local'' equations approaches. 
Note that quantities $j_o$, $\delta_o$, $V_R$, $(\partial_y V_y)_o$ 
and $(\partial_y B_x)_o$ are ``local'' (i.e.~defined at the layer central 
point~$O$ and at point~$M$ in the upstream region), while quantities $L'$, 
$V_{\rm out}'$ and $B_x'$ are ``global'' (i.e.~defined at point~$O'$ in 
the downstream region).

There are a few additional equations that we need. First, we use the 
second-order Taylor expansion of $\eta j_z$ along the y-axis to
estimate the $\eta'\!j_z'$ term in the 6th line in Table~\ref{TABLE},
\beq
\eta'j_z'&\approx&\eta_oj_o+(L'^2/2)\left\{j_o(\partial_y^2\eta)_o+{}\right.
\nonumber\\
&&\left.{}+[\eta_o+j_o(\partial\eta/\partial j_z)_o](\partial_y^2j_z)_o\right\}.
\label{TAYLOR_EXPANSION}
\eeq
Note that the first-order Taylor expansion terms are zero because of the 
symmetry. Second, we define the field global scale $L$ and the resistivity
scale $l_\eta$ as~\footnote{Here we consider 
the natural case when $(\partial\eta/\partial j_z)_o\ge0$ and 
$(\partial^2\eta/\partial y^2)_o\le0$ because plasma conductivity
decreases as the current increases and we are interested in anomalous 
reconnection that is faster than the Sweet-Parker reconnection.
\label{ETA_DERIVATIVES}
} 
\beq
L^2\equiv -2B_m\left/(\partial_y^2 B_y)_m\right.,
\quad
l_\eta^2\equiv -2\eta_o\left/(\partial_y^2\eta)_o\right..
\label{L}
\eeq
Third, the y-scale of the current $j_z$, to a factor of order unity, turns 
out to be about the same as the y-scale of the outside magnetic field,
\beq
j_o^{-1}(\partial_y^2j_z)_o\approx B_m^{-1}(\partial_y^2B_y)_m=-2/L^2.
\label{J_YY}
\eeq
This result can be understood by taking the second order partial
derivative $\partial_y^2$ of the Ampere's law equation 
$j_o\approx B_m/\delta_o$ given in the first line in Table~\ref{TABLE},
while keeping $\delta_o$ constant because the partial derivative in $y$ is
to be taken at a constant value of $x=\delta_o$. The detailed 
proof of Eq.~(\ref{J_YY}) is given in the appendix~B of the MLK2005 paper.

Now we have all equations necessary to find the reconnection rate 
and all other unknown physical parameters. In the ``Global Equations'' column
in Table~\ref{TABLE} we have six equations and six unknowns, and in the
``Local Equations'' column we have five equations and five unknowns. 
Using the ``local'' equations and 
equations~(\ref{PRESSURE_TERM}),~(\ref{L}) and~(\ref{J_YY}),
we obtain the following approximate algebraic equation for the z-current 
$j_o$ at the reconnection layer central point~$O$: 
\beq
3+\frac{j_o}{\eta_o}\left(\frac{\partial\eta}{\partial j_z}\right)_{\!o}
+\frac{L^2}{l_\eta^2}\approx\frac{\eta_o^2j_o^4L^2}{V_A^2B_m^4},
\label{RATE}
\eeq
where resistivity $\eta_o\equiv\eta(j_z=j_o,x=0,y=0)$ is a function of 
$j_o$. The ``global'' equations give a similar result with $3$ replaced 
by $1$ in Eq.~(\ref{RATE}), which is a less accurate result due to 
additional approximations made in Eq.~(\ref{TAYLOR_EXPANSION}) and in 
the formula $V_R'\approx V_R$ in the fourth line of Table~\ref{TABLE}. 

Given the resistivity function $\eta=\eta(j_z,x,y)$, as well as the 
magnetic field $B_m$ and its global scale 
$L\equiv -2B_m/(\partial_y^2 B_y)_m$ both calculated at point~$M$ 
in the upstream region, we can solve the algebraic 
Eq.~(\ref{RATE}) for the current $j_o$.  Once $j_o$ is 
calculated, we can easily find the reconnection rate, which is the rate 
of destruction of magnetic flux at point~$O$ and is equal to the 
electric field z-component $E_z(t)=\eta_oj_o$. We can also 
find all the other physical parameters, by using the equations in 
Table~\ref{TABLE},
\beq
\delta_o &\approx& B_m/j_o,
\label{DELTA}
\\
V_R &\approx& \eta_oj_o/B_m\ll V_A,
\label{V_R}
\\
(\partial_y V_y)_o &\approx& \eta_oj_o^2/B_m^2\approx V_R/\delta_o,
\label{V_y_y}
\\
(\partial_y B_x)_o &\approx& j_o\left(V_R^2/V_A^2
-2B_m^2/j_o^2L^2\right),
\label{B_x_y}
\\
B'_x &\approx& V_R\sqrt{\rho}\approx B_m(V_R/V_A),
\qquad
\label{B_x}
\\
V'_{\rm out} &\approx& V_A,
\label{V_OUT}
\\
L' &\approx& \delta_o\frac{V_A}{V_R}
\approx\frac{V_A}{(\partial_y V_y)_o}
\approx\frac{B'_x}{(\partial_y B_x)_o}.
\label{L'}
\eeq
The last approximate equality in Eq.~(\ref{L'}) is valid because the 
second term on the right-hand-side of Eq.~(\ref{B_x_y}) can be neglected.
This is because $3B_m^2/j_o^2L^2\le V_R^2/V_A^2$, which follows 
from Eqs.~(\ref{V_R}) and~(\ref{RATE}) (also refer to 
footnote~$[28]$).

Equations~(\ref{RATE})-(\ref{L'}) are very general results for
quasi-stationary magnetic reconnection with no assumptions about 
the functional form of resistivity and the guide field value. Now 
let us look at the three terms on the left-hand side of Eq.~(\ref{RATE}). 
When resistivity is constant, the only term left is the first term. 
The second term is clearly related to the dependence of anomalous 
resistivity on the current. The third term becomes important when 
resistivity is {\it ad hoc} localized in space. As a result, in the 
end of this section we consider three special cases of magnetic 
reconnection, in which Eqs.~(\ref{RATE})-(\ref{L'}) reduce to simpler 
formulas. These three cases correspond to domination of the first, 
second and third terms respectively on the left-hand-side of 
Eq.~(\ref{RATE}), and they are as follows. 

The first case is the quasi-uniform resistivity case,
\beq
\begin{array}{l}
\mbox{if~~$1\gg\max[(j_o/\eta_o)(\partial\eta/\partial j_z)_o,L^2/l_\eta^2]$,~~then}
\\
\quad j_o\approx(B_m/L)S^{1/2}, \quad\mbox{where~~$S\equiv V_AL/\eta_o$},
\\
\quad V_R/V_A\approx S^{-1/2}, \quad \delta_o\approx LS^{-1/2}, \quad L'\approx L.
\end{array}
\label{SP_RECONNECTION}
\eeq
Here we introduce the Lundquist number $S_o\equiv V_AL/\eta_o$ and assume 
for our estimates that $3^{1/4}\approx1$. Equations~(\ref{SP_RECONNECTION}) 
are the familiar Sweet-Parker results~\cite{sweet_1958,parker_1963}. Thus, if
resistivity is quasi-uniform or uniform, then reconnection is Sweet-Parker.

The second case is the Petschek-Kulsrud reconnection,
\beq
\begin{array}{l}
\mbox{if~~$(j_o/\eta_o)(\partial\eta/\partial j_z)_o\gg\max[1,L^2/l_\eta^2]$,~~then}
\\
\quad V_R/V_A \approx \delta_o/L'\approx
\left[(B_m/V_AL^2)(\partial\eta/\partial j_z)_o\right]^{1/3},
\\
\quad L'\approx L\left[(j_o/\eta_o)
(\partial\eta/\partial j_z)_o\right]^{-1/2}\ll L.
\end{array}
\label{PK_RECONNECTION}
\eeq
This is the case of fast reconnection with Petschek geometry and shocks.
Equations~(\ref{PK_RECONNECTION}) were first derived by 
Kulsrud~\cite{kulsrud_2001}. This is the reason
we call this case Petschek-Kulsrud reconnection.

The third case is the case of reconnection with spatially localized
resistivity,
\beq
\begin{array}{l}
\mbox{if~~$L^2/l_\eta^2\gg \max[1,(j_o/\eta_o)(\partial\eta/\partial j_z)_o]$,~~then}
\\
\quad j_o\approx(B_m/l_\eta)S_\eta^{1/2}, 
\quad\mbox{where~~$S_\eta\equiv V_Al_\eta/\eta_o$},
\\
\quad V_R/V_A\approx S_\eta^{-1/2}, \quad \delta_o\approx l_\eta S_\eta^{-1/2}, 
\quad L'\approx l_\eta\ll L.
\end{array}
\label{LOCALIZED_RECONNECTION}
\eeq
Here we introduce the effective Lundquist number 
$S_\eta\equiv V_Al_\eta/\eta_o$ that is based on the resistivity
scale $l_\eta$ given by Eq.~(\ref{L}). 
Equations~(\ref{LOCALIZED_RECONNECTION}) are the same as the Sweet-Parker 
equations~(\ref{SP_RECONNECTION}) with the global field scale $L$ replaced 
by the resistivity scale $l_\eta$. When resistivity is strongly localized, 
$l_\eta\ll L$, the reconnection becomes fast yielding Petschek geometry and 
shocks.
 
We postpone the detailed analysis and discussion of our theoretical results
until the last section of the paper.


\section{Reconnection with spatially localized resistivity}
\label{LOCALIZED_RESISTIVITY}

In this section we compare our theoretical results to the results of recent 
simulations of reconnection with spatially nonuniform resistivity by Baty, 
Priest and Forbes (2006)~\cite{baty_2006}. They took the resistivity as
\beq
\eta(x,y)=\eta_1+(\eta_o-\eta_1)\exp\left[-(x/l_x)^2-(y/l_y)^2\right]
\label{Baty_ETA}
\eeq
and considered different values of parameters $\eta_o$, $\eta_1$, $l_x$ and
$l_y$. They found that the reconnection rate does not depend on the value of 
$l_x$. This is in agreement with our Eq.~(\ref{RATE}), which includes
only $\partial_y^2\eta(x,y)$ derivative via resistivity scale $l_\eta$
given by Eq.~(\ref{L}). As for the dependence on the
other parameters, in all their simulation runs Baty~{\it et~al$.$} found 
the Petschek geometrical configuration with shocks and reconnection rate 
faster than the Sweet-Parker rate. They paid special attention to the case 
when $\eta_o-\eta_1$ in Eq.~(\ref{Baty_ETA}) is small and the resistivity is 
weakly nonuniform. They called this case a ``quasi-uniform resistivity'' 
case, and observed the Petschek solution in this case as well. At the same 
time, we theoretically derived the Sweet-Parker solution for the 
quasi-uniform resistivity case, given by Eqs.~(\ref{SP_RECONNECTION}). 
Thus, there is a disagreement between our theoretical results and the 
claims by Baty~{\it et~al$.$}, which needs to be addressed.

The reason for this disagreement is that the resistivity used by Baty~{\it et~al$.$} 
was not actually quasi-uniform in {\it all} their simulation runs. In fact, 
when resistivity depends only on coordinates, the condition of a 
quasi-uniform resistivity is $1\gg L^2/l_\eta^2$, refer to 
Eqs.~(\ref{SP_RECONNECTION}). In other words, the resistivity localization 
scale $l_\eta$ must be much larger than the field global scale $L$. In 
their most uniform resistivity simulation run Baty~{\it et~al$.$} used 
$\eta_o=10^{-4}$, $\eta_1=9.3\times10^{-5}$, $l_y=0.1$ and $L\approx1$. 
According to Eq.~(\ref{L}), in this case 
$l_\eta=l_y\sqrt{\eta_o/(\eta_o-\eta_1)}=0.378$, which is considerably 
smaller than $L\approx 1$. Thus, while the resistivity function 
$\eta(x,y)=9.3\times10^{-5}+7\times10^{-6}\exp[-(y/0.1)^2-(x/l_x)^2]$
would indeed appear close to uniform when its graph is plotted, its second
derivative $\partial_y^2\eta$ is large due to small value of $l_y=0.1$.
As a result, this resistivity should be viewed as far from uniform and as 
rather well localized, $L^2/l_\eta^2\approx7$. In their other 
simulation runs Baty~{\it et~al$.$} used even stronger localized resistivity 
with larger values of $\eta_o-\eta_1$ in Eq.~(\ref{Baty_ETA}). In fact, 
they were not able to run any simulations with strictly uniform resistivity 
$\eta={\rm const}$ because of an instability of the reconnection 
layer. The reason for this instability possibly lies in 
the boundary conditions that Baty~{\it et~al$.$} used, which might conflict 
with our Eq.~(\ref{RATE}). The latter must be satisfied for quasi-stationary 
reconnection.

We would like to quantitatively compare our theoretical results to the results 
obtained by Baty~{\it et~al$.$} in their most uniform resistivity simulation run. 
They had $l_\eta=0.378$, $L\approx 1$, $\eta_o=10^{-4}$, $V_A\approx 1$, 
$B_m\approx 0.9$ and they found $j_o\approx 120$ for the current at the 
reconnection layer central point~O~\cite{baty_2006,baty_2006p}. Our 
approximate Eqs.~(\ref{LOCALIZED_RECONNECTION}), derived for the localized 
resistivity case, give $j_o\approx(B_m/l_\eta)(V_Al_\eta/\eta_o)^{1/2}=146$, 
which is close to $j_o\approx 120$ observed in the simulation. The small 
disagreement could be due to plasma compressibility, or due to the finite 
Lundquist number used in the simulation and due to the approximate nature 
of our theoretical model.

We conclude that our theoretical model for magnetic reconnection is
in an agreement with the simulations by Baty~{\it et~al$.$}~\cite{baty_2006}.
Our model is in reasonable agreement with several other previous 
numerical simulations of reconnection with spatially nonuniform
resistivity~\cite{ugai_1977,tsuda_1977,scholer_1989,biskamp_2001}.


\section{Discussion}
\label{DISCUSSION}

Let us now discuss our major results.

First, equation~(\ref{RATE}) that determines the reconnection current 
$j_o$ and the quasi-stationary reconnection rate $\eta_oj_o$ is derived 
by using the ``local'' equations theoretical approach (see the 
last column of Table~\ref{TABLE}). This approach involves only ``local'' 
quantities and equations that are defined on the interval $OM$ across 
the reconnection layer, refer to Fig.~\ref{FIGURE_LAYER}. Thus, the
quasi-stationary reconnection rate in a thin two-dimensional layer is 
determined locally, in the layer central point~$O$ and in the upstream 
region outside the layer at point~$M$. In other words, the rate is 
{\it fully} determined by a particular functional form of anomalous 
resistivity $\eta(j_z,x,y)$ and by the local configuration of the 
magnetic field in the upstream region. The later determines 
field $B_m$ and its scale $L\!\equiv\![-2B_m/(\partial_y^2 B_y)_m]^{1/2}$ 
at point~$M$. As a result, the global properties of the 
reconnection layer (e.g.~its length $L'$, the plasma outflow velocity 
$V_{\rm out}'$ and the presence or absence of Petschek shocks) do not 
directly matter for the quasi-stationary reconnection rate. This is a 
very important result because the reconnection layer global geometry 
can be very complicated (e.g.~in turbulent plasmas). Of course, there 
exists an indirect dependence of the reconnection rate on the field 
global properties because the local field configuration in the upstream 
region is determined by the field global configuration. 
The above statements are also true for all other ``local'' parameters -- 
for the layer thickness $\delta_o$ and for the reconnection velocity 
$V_R$, refer to Eqs.~(\ref{DELTA}) and~(\ref{V_R}). 

Second, the reconnection rate can also be estimated by using the 
``global'' equations theoretical approach, in which the whole 
reconnection layer is considered, including the downstream region at 
point~$O'$ (refer to Fig.~\ref{FIGURE_LAYER}). The equations used in 
this approach are presented in the second column of Table~\ref{TABLE}. 
We would like to point out that two of these equation play the key role 
in correct determination of the reconnection layer length $L'$ and geometry. 
The first key equation is the jump condition on the Petschek shocks, 
$B_x'/\sqrt{\rho}\approx V_R'\approx V_R$, whose importance was first 
pointed out by Kulsrud~\cite{kulsrud_2001}. The second key equation is 
the constancy of the electric field z-component along the reconnection 
layer, $\eta_oj_o=\eta'j_z'+V_{\rm out}'B_x'$, which has been 
overlooked in the previous theoretical models (e.g.~in the Petschek 
model).

Third, for the case of a strong dependence of resistivity on the 
current, i.e.~when
$(j_o/\eta_o)(\partial\eta/\partial j_z)_o\gg\max[1,L^2/l_\eta^2]$, 
our results, given by Eqs.~(\ref{PK_RECONNECTION}), coincide
with the results obtained by Kulsrud~\cite{kulsrud_2001}. Thus,
both the ``local'' and ``global'' theoretical approaches confirm Kulsrud's 
results and ideas, contrary to the doubts raised in paper by 
Baty {\it et~al$.$}~\cite{baty_2006}.

Fourth, we found that in the case of uniform or quasi-uniform 
resistivity the magnetic reconnection rate is the slow Sweet-Parker 
rate and not the fast Petschek rate, see Eqs.~(\ref{SP_RECONNECTION}).
This theoretical result follows from rigorous analytical derivations 
and agrees with numerical simulations. At the same time it contradicts 
the original Petschek theoretical model. Let us consider both the 
``global'' and ``local'' analytical approaches in the case of 
a strictly uniform resistivity $\eta={\rm const}=\eta_o$, and let 
us explain why the Petschek reconnection layer geometry is not 
realized in this case. We take the ``global'' equations approach first. 
In the case of constant resistivity Eqs.~(\ref{TAYLOR_EXPANSION}) 
and~(\ref{J_YY}) result in equation 
$(\eta_oj_o-\eta'j_z')/\eta_oj_o\approx (L'/L)^2$ for the fractional
drop of the $\eta j_z$ term along the reconnection layer.
On the other hand, the constancy of the electric field z-component 
along the reconnection layer implies that this fractional drop is
$(\eta_oj_o-\eta'j_z')/\eta_oj_o\approx V_{\rm out}'B_x'/\eta_oj_o\approx 1$,
where we use the ``global'' equations on lines $3$ to~$6$ in the 
second column of Table~\ref{TABLE}. These two equations agree only 
if $L'\approx L$, which means that the geometry of the reconnection 
layer is Sweet-Parker and not Petschek. Next, we take the ``local'' 
equations approach. In this ``local'' approach we prefer not to consider 
the reconnection layer geometry and any ``global'' parameters, such as 
the layer length $L'$, for the calculation of the reconnection rate. 
Instead we argue as follows. Refer to the ``local'' equations in the 
last column of Table~\ref{TABLE}. Equations on lines~$1$, $2$ and~$5$ 
result in an estimate of the plasma outflow velocity derivative
$(\partial_y V_y)_o\approx\eta_oj_o^2/B_m^2$, see Eq.~(\ref{V_y_y}).
Plasma acceleration equation on line~$3$ results in the upper estimate 
for the $B_x$ field derivative, 
$(\partial_y B_x)_o\le\rho{(\partial_y V_y)_o}^2/j_o$. As a result, we 
can find the upper estimate for the reconnection current $j_o$ from 
the ``local'' equation on line~$6$, which is the condition of constancy
of the electric field z-component along the reconnection layer. If 
resistivity is uniform, this estimate turns out to be the 
Sweet-Parker value, $j_o\approx (B_m/L)(V_AL/\eta_o)^{-1/2}$, as 
given by Eq.~(\ref{SP_RECONNECTION}). In addition, if one estimates 
the reconnection layer length as  
$L'\sim B_x'/(\partial_y B_x)_o\sim V_A/(\partial_y V_y)_o$, he/she
would again recover the Sweet-Parker result $L'\approx L$~\cite{malyshkin_2005}.
We conclude that in the case of constant resistivity both the ``global'' 
and ``local'' equations approaches consistently lead to the Sweet-Parker 
reconnection rate and the Sweet-Parker geometry of the reconnection 
layer ($L'\approx L$).

Finally, let us point out that whether reconnection is unforced (free) 
or forced does not matter for our results and conclusions. Indeed, on 
one hand, in the case of unforced reconnection one solves 
Eqs.~(\ref{RATE}) and~(\ref{V_R}) for the current $j_o$ and for the 
reconnection velocity $V_R$. The solution will depend on the magnetic
field $B_m$ in the upstream region at point~$M$. On the other 
hand, in the case of forced magnetic reconnection the reconnection 
velocity $V_R$ is prescribed and fixed. In this case the field 
$B_m$ in the upstream region should be treated as an unknown parameter, 
and Eqs.~(\ref{RATE}) and~(\ref{V_R}) are to be solved together in
order to find the correct quasi-stationary values of $j_o$ and
$B_m$. In other words, in the forced reconnection case an initially
weak outside magnetic field $B_m$ gets piled up to higher values 
until the resulting current $j_o$ in the reconnection layer
becomes large enough to be able to match the prescribed velocity 
$V_R$ of magnetic flux and energy supply in the upstream region.


\begin{acknowledgments}
It is our pleasure to thank Hantao Ji, Dmitri Uzdensky and 
Masaaki Yamada for stimulating discussions and valuable comments.
This work was supported by the Center for Magnetic Self-Organization
(CMSO) grant. L.~M.'s research was also partly supported by 
NASA Grant No.~NNG04GD90G. L.~M.~would like to acknowledge hospitality 
of the Aspen Center for Physics, where the results of this work were 
presented for the first time.

\end{acknowledgments}


\end{document}